\documentclass[prl,twocolumn,preprintnumbers,amsmath,amssymb,showpacs,superscriptaddress]{revtex4} 

\usepackage{graphicx}
\usepackage{amsmath,amssymb,amsfonts}
\usepackage{bm}
\usepackage[T1]{fontenc}
\usepackage[english]{babel}
\usepackage[latin1]{inputenc}
\usepackage[usenames,dvipsnames]{color}
\usepackage{hyperref}
\hypersetup{
  colorlinks=true,
 citecolor=blue}
  
\usepackage{graphicx}
\usepackage{dcolumn}
\usepackage{latexsym,amsmath,amssymb,amsfonts,amsthm,enumerate}

\newcommand{\nn}{\nonumber}

\renewcommand{\(}{\left (}
\renewcommand{\)}{\right )}
\renewcommand{\[}{\left [}
\renewcommand{\]}{\right ]}
\newcommand{\la}{\langle}
\newcommand{\ra}{\rangle}
\renewcommand{\o}{\omega}

\newcommand{\dg}{\dagger}

\renewcommand{\|}{\nn \\}

\newcommand{\eq}[2][]{
\begin{equation}
#2 \label{#1} 
\end{equation}} 

\renewcommand{\!}{\nn \\ & }
\newcommand{\eqa}[2][]{
\begin{align}
 #2 \label{#1}
\end{align}}

\begin{document}

\title{Quantum Quench of the trap frequency in the harmonic Calogero model}

\author{M. A. Rajabpour}
\affiliation{Instituto de F\'{\i}sica de S\~{a}o Carlos, Universidade de S\~{a}o Paulo, Caixa Postal 369, 13560-590, S\~{a}o Carlos, SP, Brazil}
\author{S. Sotiriadis}
\affiliation{Dipartimento di Fisica dell'Universit\`a di Pisa and INFN, 56127 Pisa, Italy}

\date{\today}

\begin{abstract}
We consider a quantum quench of the trap frequency in a system of bosons interacting through an inverse-square potential and confined in a harmonic trap (the harmonic Calogero model). We determine exactly the initial state in terms of the post-quench eigenstates and derive the time evolution of simple physical observables. Since this model possesses an infinite set of integrals of motion that allow its exact solution, a generalised Gibbs ensemble (GGE), i.e. a statistical ensemble that takes into account the conservation of all integrals of motion, can be proposed in order to describe the values of local physical observables long after the quench. Even though, due to the presence of the trap, physical observables do not exhibit equilibration but periodic evolution, such a GGE may still describe correctly their time averaged values. We check this analytically for the local boson density and find that the GGE conjecture is indeed valid, in the thermodynamic limit.
\end{abstract}


\maketitle

\emph{Introduction}. --  Sparked by experimental findings in the field of ultracold atoms out-of-equilibrium \cite{uc,kww-06,tc-07,tetal-11,cetal-12,getal-11,shr-12,rsb-13}, questions about the time evolution of quantum systems have become the subject of intense study. Without doubt, the investigation of whether \emph{thermalization} or some more general \emph{equilibration} occurs when starting from an out-of-equilibrium initial state, has been established as the main objective (\cite{revq} for a review). A common protocol for the preparation of the initial state is a \emph{quantum quench}, i.e. an instantaneous change of the parameters of the Hamiltonian of the system so that the initial state is the ground state of the pre-quench Hamiltonian. In integrable systems, i.e. 1d systems possessing an infinite number of local integrals of motion (IoM), the evolution is constrained by the extra conservation laws and thermalization is prevented. However a generalized relaxation incorporating the extra constraints is still possible and in fact it has been demonstrated that such a \emph{generalized Gibbs ensemble} (GGE) describes the large time values of \emph{local} physical observables in various settings \cite{cc-06,gg,c-06,c-06b,cdeo-08,bs-08,r-09,CEF,f-13,eef-12,se-12,ccss-11,rs-12,bdkm-11,fm-10,mc-12a,ce-13,fe-13,CSC13a}. 

Most analytical demonstrations of the validity of the GGE refer to non-interacting systems or systems that can be mapped into non-interacting ones via some suitable nontrivial transformation \cite{CEF,c-06,c-06b} with the few exceptions that refer to genuine interacting systems (i.e. described by nontrivial scattering phase shifts), restricting to special classes of initial states \cite{fm-10}. For such genuinely interacting systems even the derivation of GGE predictions is a difficult task that has been accomplished for some models only recently \cite{gge-new}. A common obstacle in the study of quantum quenches in these systems is the derivation of the expansion of the initial state on the eigenstates of the post-quench Hamiltonian \cite{sfm-12}. 
Despite the technical difficulties, testing the GGE conjecture for a quantum quench in a genuinely interacting system is necessary, especially since in a non-interacting system the verification is somewhat expected because all observables can typically be derived by the IoM themselves. Moreover, unlike the physics of ground state or thermal equilibrium of an interacting system which is governed by its low energy properties that usually can be effectively described by a non-interacting model, this is certainly not true after a quantum quench, as far as the question of equilibration is concerned, since the high energy excitations play in general a significant role.

In this letter we consider the harmonic Calogero model (HCM) \cite{Calo,Suth}, i.e. a system of harmonically trapped particles interacting via an inverse-square potential. We restrict our attention to the case of bosonic elementary particles. The Calogero model in infinite space and its variants in a harmonic trap or on a circle (Calogero-Moser-Sutherland models \cite{Calo,MC75,Suth}) possess an infinite set of IoM and, even though genuinely interacting, are well-known for exhibiting effectively free behaviour with generalized particle statistics \cite{anyons}. In the harmonic trap, the spectrum is equidistant and the Hamiltonian can be diagonalised by means of creation-annihilation operators satisfying generalised canonical commutation relations (CCR) \cite{Poly92}. This allows the exact derivation of the initial state after a quench of the trap frequency and the study of the subsequent time evolution. 

The existence of an infinite set of IoM suggests that a GGE may be applicable. However, due to the presence of the trap, 
the excitation levels are equidistant and the periodicity of the evolution does not allow equilibration of observables, even in the thermodynamic limit \cite{CSC13b}. In this case one may still apply a weak version of the GGE conjecture (cf. \cite{bch-11}) stating that it describes the \emph{time averages} of local observables. Despite the fact that in a trapped system the \emph{locality} of the IoM (which is a characteristic feature of quantum integrability \cite{CM11} and is typically considered as a condition for the validity of the GGE \cite{fe-13}) is lost or ambiguous, this conjecture has been recently verified in the case of the essentially non-interacting Tonks-Girardeau gas in a quenched trap \cite{CSC13b}.

We derive analytically the GGE predictions for the time averaged values of the local density of particles and compare them with their actual values. We show that the GGE predicts \emph{correctly} the actual values, when the thermodynamic limit is taken into account.

\emph{The model}. -- The HCM is described by the Hamiltonian
\eq{ H= \sum_{i=1}^N \frac12 p_i^2 + \sum_{i=1}^N \frac12\omega^2 x^2_i + g \sum_{i<j} \frac1{(x_i-x_j)^2} }
where $N$ is the number of particles and $\omega$ the trap frequency. The interaction constant is more conveniently parametrized as $g=\ell (\ell-1)$. Its ground state is given by the wavefunction
\eq[gsw]{ \psi_{gs}(\o;\{x_i\}) = \mathcal{N}_{N,\ell} \; \o^{N[1+\ell (N-1)]/4 } e^{-\frac12 \omega \sum_i x_i^2} \Delta(\{x_i\})^\ell  }
where $\Delta(\{x_i\}) \equiv \prod_{j<k} (x_j-x_k)$ is the Vandermonde determinant and $\mathcal{N}_{N,\ell}$ a normalisation constant. The ground state energy is $ E_{gs} = \frac12 N \o [ 1 + \ell {(N-1)} ] $. 
The local density in the ground state $ \rho_{gs}(\o;x) \equiv \int  |\psi_{gs}(\o;x,\{x_i\}_{i\neq 1})|^2 \prod_{i\neq 1} dx_i$ is given for large $N$ by the \emph{Wigner semicircle} distribution \cite{Suth71}
\eq[dens]{  \varrho_{gs}(\o;x) = \frac{2N}{\pi x_0} \sqrt{ 1- \(\frac{x}{x_0}\)^2 } \quad \text{ if  } |x| \le x_0 \equiv \sqrt{\frac{2N\ell}{\o}}}
and zero otherwise. Notice that the density is independent of $\ell$, except through the cloud radius $x_0$. 

One method to study this model and derive its energy eigenstates is by using the \emph{Dunkl} or \emph{exchange operator} formalism \cite{Dunkl89,Poly92}, in which the interaction term is written as $\sum_{i<j} {\ell (\ell-M_{ij})/(x_i-x_j)^2}$ where $M_{ij}$ is the so-called exchange operator that permutes the coordinates of two particles $i$ and $j$. Assuming that all particles are bosonic, the introduction of the exchange operator does not alter the physics. $M_{ij}$ satisfies the properties $M_{ij}=M_{ij}^{-1}=M_{ij}^\dg=M_{ji}$, $[M_{ij},M_{kl}] = 0$ if $i,j,k,l$ distinct, $M_{ij}M_{jk}=M_{ik}M_{ij}$ if $i,j,k$ distinct and $M_{ij}A_k=A_k M_{ij}$ if $i,j,k$ distinct, $ M_{ij}A_i=A_j M_{ij} $ for any operator $A_i$. Now defining creation-annihilation operators
\eqa[xp]{a_i = \sqrt{{\o}/{2}} \(x_i + i\pi_i/\o \) \| 
a_i^\dg = \sqrt{{\o}/{2}} \(x_i - i\pi_i/\o \) } 
where $\pi_i \equiv p_i + \sum_{j\neq i} \frac{i\ell}{x_{ij}} M_{ij} $, 
we find that they satisfy generalized canonical commutation relations (CCR)
\eqa[ccr]{ [a_i,a_j] & =[a_i^\dg,a_j^\dg]=0 \| 
[a_i,a_i^\dg] & = 1+\ell {\textstyle \, \sum_{j\neq i} M_{ij}} \| 
[a_i,a_j^\dg] & = -\ell M_{ij} \quad \text{ if } i\neq j }
Using the above definitions, the Hamiltonian can be cast in the diagonal form 
${ H = \sum_i \frac12 \o (a_i^\dagger a_i + a_i a_i^\dagger ) }$. 

There are several alternative but equivalent ways to write the set of commuting IoM for this model. A common choice is \cite{Poly92}
\eq[iom1]{I_s= \sum_{i=1}^N (a_i^{\dg} a_i)^s }
with $s=1,2,...,N$. These constitute a minimal complete set of IoM since all higher ones are algebraically dependent on the lowest $N$. Another choice is \cite{jap98} ${J_i= a_i^{\dg} a_i + \ell \sum_{j=1}^{i-1} (M_{ij} - 1) }$. 
It should be emphasised that \emph{none} of these sets of IoM are local \emph{nor} manifestly equivalent to local ones, as typically required by the GGE conjecture \cite{fe-13}. To the best of our knowledge, no definition of a set of local IoM exists for the HCM.

\emph{Trap quench and the initial state}. -- Now we consider a quantum quench of the trap frequency from $\o_0$ to $\o$ and wish to write the initial ground state $|\Psi\ra$ in terms of the post-quench eigenstates. From (\ref{xp}) we find that the post-quench creation-annihilation operators $a_i,a_i^\dg$ are related to the pre-quench ones $a_{0i},a_{0i}^\dg$ by a \emph{Bogoliubov transformation}, as in the non-interacting case \cite{cc-06}
\eq[bog]{ a_{0i} = \frac12 \[ \(\sqrt{\frac{\o_0}{\o}} + \sqrt{\frac{\o}{\o_0}}\) a_i + \(\sqrt{\frac{\o_0}{\o}}-\sqrt{\frac{\o}{\o_0}}\) a_i^\dg \]}
which means that $|\Psi\ra$ satisfies the equation $( a_i + \kappa a_i^\dg )|\Psi\ra=0$ where $\kappa \equiv (\o_0-\o)/(\o_0+\o)$, from which we can find its expansion on post-quench eigenstates. Remarkably, regardless the nontrivial form of the CCR (\ref{ccr}), the initial state turns out to be of the same \emph{squeezed coherent} form as in the non-interacting case \cite{SGS13}
\eq[psi]{|\Psi\ra = \mathcal{N}_\kappa \exp\Big( -\frac12 \kappa \sum_i a_i^{\dg 2} \Big)|0\ra}
where $|0\ra$ is the post-quench ground state ($a_i|0\ra=0$). Written in this form, the state can be readily evolved in time
\eq[psit]{e^{-iHt}|\Psi\ra = \mathcal{N}_\kappa \exp\Big( -\frac12 \kappa e^{-2i\o t} \sum_i a_i^{\dg 2} \Big)|0\ra.}
States of the form above are well-known as \emph{squeezed vacua}, they are produced by the action of the \emph{squeeze operator} $S(\xi) \equiv \exp\Big[ \sum_i (\xi^* a_i^{2} - \xi a_i^{\dg 2})/2 \Big]$ on the vacuum \cite{Perel86} and, equivalently, have the characteristic property of being annihilated by a \emph{squeezed annihilation operator} $S^\dg a S$, like the one of (\ref{bog}). Indeed, as shown in \cite{SM}, the state (\ref{psit}) can also be written in the form $S(\xi)|0\ra$ for $ \xi \equiv \xi(t) = \frac12 e^{-2i\o t} \log(\o_0/\o) $.

\emph{Coordinate space representation of the wavefunction}. -- Squeezed coherent states, like coherent states too, have a simple interpretation when seen as wavefunctions in coordinate (or in phase) space. Obviously by its definition, the initial state $|\Psi\ra$ corresponds in coordinate space to the wavefunction of the post-quench ground state rescaled by a factor $\o_0/\o$. An elegant way to see this from (\ref{psi}) is by using its alternative form $|\Psi\ra = S(\xi)|0\ra$ and noticing that $\xi(0)$ is real and $ \frac12 \sum_i (a^2_i - {a_i^\dg}^2) = N/2 + \sum_i x_i \partial_{x_i} $ is essentially the generator of uniform coordinate scalings \cite{SM}. 

For the evolved state (\ref{psit}) the amplitude $\kappa e^{-2i\o t} \equiv \eta(t)$ is complex and the calculation is more elaborate. The final result \cite{SM} is
\eqa[5]{ \psi& (\{x_i\},t) =  \mathcal{N}_{N,\ell} \;  \[\o \frac{{1-|\eta(t)|^2}}{(1-\eta(t))^2}\]^{N [1+\ell (N-1)]/4} \times \| 
& \times  \exp\[-\frac12 \o \( \frac{1+\eta(t)}{1-\eta(t)} \) \sum_i x_i^2 \]  \Delta(\{x_i\})^\ell  }
This expression is in agreement with and can be derived in a completely different way by using a scaling transformation to solve the problem of a system confined in a time-dependent harmonic trap \cite{Suth98,GBD10} and specialising to the quench protocol (cf. \cite{mg-05} for the Tonks-Girardeau limit).

\emph{Time evolution of observables}. -- 
Knowing the evolution of the system's state we proceed to calculate the expectation values of physical observables. Since our aim is to compare with the GGE predictions, we will focus on the density profile $\varrho(x,t) \equiv \int |\psi(x,\{x_i\}_{i\neq 1},t)|^2 \prod_{i\neq 1} dx_i$ which is a \emph{local} observable, and calculate its time average $\bar{\varrho}(x) \equiv \o/\pi \int_0^{\pi/\o} \varrho(x,t) dt $ in the thermodynamic limit. However we will also derive its moments $\la x^{2n} \ra $, i.e. the expectation value of the operators $x_i^{2n}$ of any of the particles, which are {nonlocal} observables but related to $\varrho(x,t)$ through $\la x^{2n} \ra \equiv \int x^{2n} \varrho(x,t) dx $, so that the latter can be reconstructed when all of $\la x^{2n} \ra $ are known. The thermodynamic limit for trapped systems is defined so that the system size and number of particles tend to infinity, but the density (at any point in the bulk or averaged over all space) remains finite. This requirement is fulfilled when the trap frequency (in our case both $\o_0$ and $\o$) scales like $1/N$ \cite{CSC13a,CSC13b}. Since the density does not scale with $N$ in this limit, the moments scale like ${\la x^{2n} \ra } \sim N^{2n}$ and this leading order is the only one we are interested in.

\begin{figure}[h!]
\centering
\includegraphics[width=.95\columnwidth]{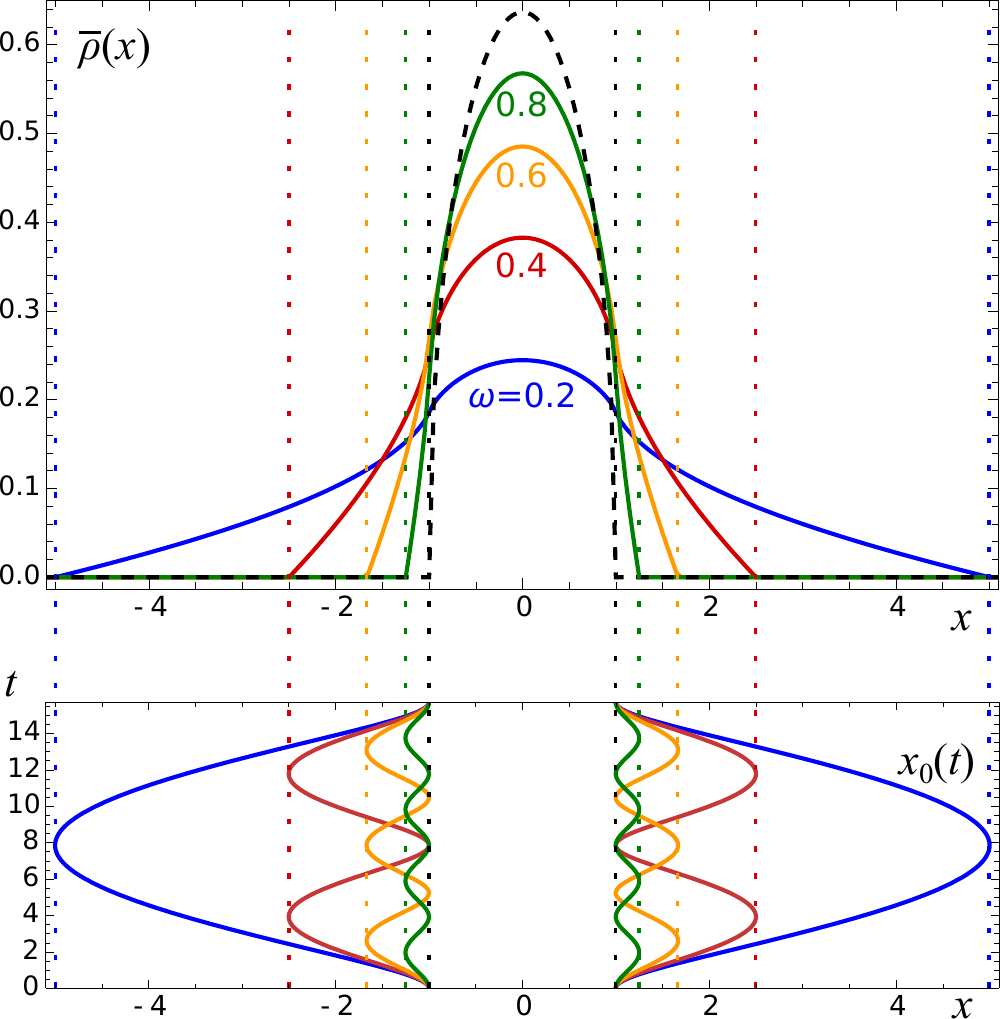}
\caption{\label{F:rho} \small \emph{Top:} Time-averaged density profile $\bar{\varrho}(x)$ (in units $N^{-1}$) as a function of the position $x$ (in units $(N\ell)^{-1/2}$) for fixed pre-quench trap frequency $\omega_0=1$ and variable post-quench trap frequency $\omega = 0.2, 0.4, 0.6, 0.8$. The dashed curve indicates the initial Wigner-semicircle density profile $\varrho_{gs}(\omega_0;x)$. \emph{Bottom:} Time evolution of the cloud edges $\pm x_0(t) = \pm\sqrt{2N\ell/\Omega(t)}$ for the corresponding values of $\omega$. The vertical dotted lines indicate the positions of the extrema of $x_0(t)$.}
\end{figure}

Since the evolved state wavefunction (\ref{5}) is a scaling transformation of that of the ground state, we can readily show that the corresponding density profile $\varrho(x,t)$ is also given by the Wigner semicircle (\ref{dens}) after replacing $\o$ in $x_0$ by
\eq[Omegat]{ \Omega(t) \equiv 2\text{Re} \( \frac{1+\eta(t)}{1-\eta(t)} \) \o =  \frac{2 \o (1-\kappa^2)}{1+\kappa^2 -2 \kappa \cos 2 \o t} } 
i.e.  $\varrho(x,t) = \varrho_{gs}(\Omega(t);x) = \varrho_{gs}(\omega;x\sqrt{\Omega(t)/\omega}) \sqrt{\Omega(t)/\omega}$ (cf. \cite{Suth98}). As expected, the density profile exhibits oscillatory (``breathing'') behaviour. 
The time averaged density profile $\bar{\varrho}(x)$ is evaluated numerically and plotted in Fig.~\ref{F:rho} for several values of $\kappa$. From the above results it is easy to calculate explicitly the moments of $\varrho(x,t)$ and their time averaged values \cite{SM}.

\emph{Generalised Gibbs Ensemble}. --  Having found the time averages of observables we can now compare them with the corresponding GGE predictions. It is convenient to use as test observables the moments $\overline{\la x^{2n} \ra} $ from which the local observable $\bar\rho(x)$ can be reconstructed.

To construct the GGE density matrix we use the IoM $I_s$ given by (\ref{iom1}). This choice of IoM corresponds to the ones used in the exact solution of the classical version of the HCM \cite{ciom}. Besides, it is a natural generalisation \cite{SM} of the set considered in \cite{CSC13a} which is the only other existing demonstration of the GGE conjecture in a trapped system. The corresponding GGE density matrix is
\eq[rhogge]{ \rho_{\text{GGE}} =  \frac{e^{-\sum_{s=1}^N \beta_s I_s}}{\mathcal{Z}} = \frac{1}{\mathcal{Z}} \exp\({-\sum_{s=1}^N \sum_{i=1}^N \beta_s (a^\dg_i a_i)^s }\) }
where $\mathcal{Z} \equiv \text{Tr} \; \exp({-\sum_{s=1}^N \beta_s I_s}) $ and the Lagrange multipliers $\beta_s$ are determined by the condition that the GGE values of the IoM are equal to their values in the initial state, i.e. $\la \Psi | I_s | \Psi\ra = \text{Tr} \( \rho_{\text{GGE}} \, I_s \) = -{\partial}(\log \mathcal{Z})/{\partial \beta_s} $. 
However deriving the $\beta_s$ is not a necessary step, since it is possible to directly express the moments in terms of the IoM and their products, which can be calculated in the GGE without explicit knowledge of the $\beta_s$.

Indeed, by expanding the operator $x_i^{2n} = ( a_i + a_i^\dg )^{2n}/{(2\o)^n}$ and evolving in time (in the Heisenberg picture) 
we realise that \emph{only} terms with equal number of $a_i$ and $a_i^\dg$'s contribute to the time averaged expectation values, since only those are unaccompanied by oscillatory phase factors. But these terms can be recast as powers and products of the operators $ a_i^\dg a_i$ using the commutation relations (\ref{ccr}) so that in the end we find \cite{SM}
\eqa[taevo]{ & \overline{\la x^{2n} \ra} = \frac1N \frac1{(2\o)^n} \Bigg( \sum_{m=0}^n d_{n,m}  {\la I_m \ra}_0 + \|
& \quad + \sum_{m=2}^{n-1} \sum_{\text{all possible} \atop \text{partitions of } m} 
\,  e_{n,m,\{k_s\}}  {\left\la \textstyle\prod^{r}_{s=1}I_{k_s} \right\ra}_0 \Bigg) }
for suitable coefficients $d_{n,m}$ and $e_{n,m,\{k_s\}}$, where $\{k_s\}$ is a partition of $m$ in terms of $r$ numbers, i.e. $\sum_{s=1}^r k_s = m $. Obviously the expectation values of the IoM and their products are calculated in the initial state, as the index ``0'' indicates. Note that the presence of correlators of IoM products is an effect of the non-trivial form of the CCR (\ref{ccr}) for $\ell\neq 0$.

On the other hand, using the same expansion to calculate the GGE averages 
$ \la x_i^{2n} \ra_{\text{GGE}} = \text{Tr} \( x_i^{2n}  \rho_{\text{GGE}} \) $
we find that it is exactly those \emph{same} terms with equal number of $a_i$ and $a_i^\dg$'s that contribute, since only those have non-zero traces in the basis of post-quench eigenstates. Therefore $ \la x^{2n} \ra_{\text{GGE}} $ is given by the RHS of (\ref{taevo}) but with the expectation values of the IoM and their products calculated now in the GGE instead of the initial state. 

The values of the IoM in the initial state and in the GGE are by definition equal to each other ${\la I_m \ra}_0 = {\la I_m \ra}_{GGE} $. On the other hand, their products are uncorrelated in the GGE in the thermodynamic limit, i.e. $ {\left\la \textstyle\prod^{r}_{s=1}I_{k_s} \right\ra}_{GGE} = \textstyle\prod^{r}_{s=1} {\left\la I_{k_s} \right\ra}_{GGE} $. This statement is based on the fundamental concept of statistical physics that, in the thermodynamic limit, the expectation value of some operator in a statistical ensemble equals its value on a single eigenstate (microstate) that is \emph{representative} of the ensemble \cite{mc-12b,ce-13,yy69} (the relative fluctuations about these values scale to zero with $N$). Since the IoM are diagonal on the eigenstates, their products factorise. Therefore what determines the validity of the GGE conjecture is whether this property holds also in the initial state. This turns out to be true, when we consider the thermodynamic limit. To see this \cite{SM}, we first have to calculate the values of the IoM and their products in the initial state by using the inverse of the Bogoliubov transformation (\ref{bog}) and normal-ordering according to the CCR (\ref{ccr}). Keeping only terms that contribute to leading order in $N$ we verify that for an arbitrary product of IoM we have $ {\left\la \textstyle\prod^{r}_{s=1}I_{k_s} \right\ra}_0 =  \textstyle\prod^{r}_{s=1} {\left\la I_{k_s} \right\ra}_0 $ \cite{SM}. According to the above, it finally follows that the GGE predicts \emph{correctly} the values of $\overline{\la x^{2n} \ra}$ and therefore of $\bar{\rho}(x)$. 

\emph{Conclusions.} -- The above analysis shows that the GGE conjecture is valid for a quantum quench of the trap frequency in the HCM. It should be stressed that the thermodynamic limit was an essential step in the course of this verification. Finite size corrections coming from lower order contributions in $N$ would spoil the crucial property of the IoM products to have uncorrelated values in the initial state. 

The same could happen if a different initial state was considered, since this property is not necessarily true for any initial state. In this case the GGE would not predict correctly the $\overline{\la x^{2n} \ra}$ since it would miss information about the initial correlations between the IoM. In order to correct the GGE, one would have to generalise it so that (\ref{rhogge}) includes not only the IoM but also all products of them, with the values of the corresponding additional $\beta$'s fixed independently from the initial condition \cite{gp08,KE08,c-06b}. Also note that the above refer to the general case $\ell \neq 0$; in the trivial case of noninteracting bosons, i.e. $\ell = 0$, the GGE is correct anyway, since all terms involving products of IoM in (\ref{taevo}) disappear and the equality of the values of the IoM in the initial state and in the GGE is sufficient to ensure its validity.

Another remark based on (\ref{taevo}) is that in order to correctly predict $\overline{\la x^{2n} \ra }$ it is sufficient to use a \emph{truncated} GGE \cite{fe-13} with only the lowest $n$ IoM fixed by the initial conditions. In particular, a measurement of $\overline{\la x^2\ra}$ \emph{only}, would lead to the misconception that the system is described by a Gibbs ensemble. 
This is not in contradiction with the GGE conjecture, since the latter refers to local observables while the moments are global ones. In order to correctly predict the time averaged local density profile $\bar{\varrho}(x)$ instead, we need all of its moments $\overline{\la x^{2n}\ra}$ and therefore the \emph{full} GGE is required, i.e. including all infinite IoM. Notice that in (\ref{rhogge}) we used only the $N$ lowest IoM: including all higher ones as well would be equivalent to including all products of the $N$ lowest ones, since the higher are algebraically dependent on the lower ones. However, as discussed above, in the thermodynamic limit the products of IoM do not need to be included in the GGE, because their values in the initial state become uncorrelated and therefore are already predicted correctly by the GGE (\ref{rhogge}).

Even though an experimental implementation of the HCM remains elusive so far, we expect that the general conclusions drawn above will serve as guidelines to experimentalists working on the verification of the GGE conjecture, since the need of a confining trap   (that breaks the integrability of other models, like the Lieb-Liniger) in experimental systems seems inevitable.

\emph{Acknowledgments}. 
We are grateful to P. Calabrese and G. Takacs for useful discussions. M.~A.~R. acknowledges FAPESP for financial support. S.~S. acknowledges the ERC for financial support under Starting Grant 279391 EDEQS.



\newpage

\onecolumngrid

\section{SUPPLEMENTAL MATERIAL}

\subsection{Squeezed states and $su(1,1)$ algebra}

In this section we will derive some properties of squeezed states, mentioned in the main text, using algebraic properties of the creation-annihilation operators. First we will show the equivalence of the two alternative forms of squeezed states and second we will derive their coordinate space representation.

\emph{1.} To see that the state (\ref{psit}) can also be written in the form $S(\xi)|0\ra$ for a suitable value $\xi$ and vice versa, one may notice that, due to the commutation relations (\ref{ccr}), the operators
\eqa{ 
K_- & \equiv \sum_i a^2_i/2 \| 
K_+ & \equiv \sum_i {a_i^\dg}^2/2 \| 
K_0 & \equiv \sum_i (a_i^\dg a_i + a_i a_i^\dg)/4 = H/(2\o) }
satisfy the $su(1,1)$ algebra: $[K_0,K_\pm]=\pm K_\pm$ and $[K_-,K_+]=2K_0$, i.e. they are generators of the $SU(1,1)$ group. This allows us to ``\emph{disentangle}'' the exponential operator $S(\xi) = \exp(\xi^* K_- - \xi K_+ )$, i.e., to write it as a product of exponential operators $S(\xi) = \exp(\alpha(\xi) K_+) \exp(\beta(\xi) K_-) \exp(\gamma(\xi) K_0)$. The values of the coefficients $\alpha, \beta$ and $\gamma$ as functions of $\xi \equiv r e^{i\phi}$ can be derived, for example, by considering a finite matrix representation of the $SU(2)$ group (whose algebra is trivially related to that of $SU(1,1)$ which, being a non-compact Lie group, has no finite representation) and simply solving a small set of equations \cite{Wilc67} 
\eqa{ \alpha(\xi) & = - \tanh r \; e^{i\phi} \|
\beta(\xi) & =  \sinh r \cosh r \; e^{-i\phi} \|
\gamma(\xi) & = - 2 \log{(\cosh r)} }
It is then trivial to see that the disentangled operator acting on $|0\ra$ yields
\eq[Aaf]{ S(\xi)|0\ra = (1-|\alpha(\xi)|^2)^{ N ( 1 + \ell {(N-1)} )/4 } \exp\Big( \alpha(\xi) K_+ \Big)|0\ra }
which is exactly (\ref{psit}) for $\xi$ given by
\eq[xi]{ \xi \equiv \xi(t) = \frac12 e^{-2i\o t} \log\(\frac{1+\kappa}{1-\kappa}\) = \frac12 e^{-2i\o t} \log(\o_0/\o) }

For the initial state wavefunction $\psi(\{x_i\})$, noticing that the amplitude $\kappa$ is real, and therefore so is $\xi(0)$, and that $ \frac12 \sum_i (a^2_i - {a_i^\dg}^2) = N/2 + \sum_i x_i \partial_{x_i} $ is essentially the generator of uniform coordinate scalings, we find from (\ref{gsw}) and the above results
\eqa[A3]{& \psi(\{x_i\}) = e^{\xi N/2} \psi_{gs}(\o;\{e^\xi x_i\}) \| 
& = \mathcal{N}_{N,\ell} \; \o^{N[1+\ell (N-1)]/4 } e^{\xi N [1+\ell (N-1)]/2 -\frac12 \o e^{2\xi} \sum_i x_i^2} \Delta(\{x_i\})^\ell \|
& = \mathcal{N}_{N,\ell} \; \o_0^{N[1+\ell (N-1)]/4 } e^{-\frac12 \o_0 \sum_i x_i^2} \Delta(\{x_i\})^\ell \|
& =  \psi_{gs}(\o_0;\{x_i\}) }
which verifies that the squeezed vacuum state (\ref{psi}) gives the correct pre-quench ground state wavefunction.

\emph{2.} We saw that the initial state $|\Psi\ra$ corresponds, in coordinate space, to the wavefunction of the post-quench ground state rescaled by a factor $\o_0/\o$. For the evolved state (\ref{psit}) the amplitude $\kappa e^{-2i\o t}/2 \equiv \eta$ is complex and the calculation of its coordinate space wavefunction is more elaborate. We first write the squeeze operator as 
\eq{ S(\xi) = e^{  \text{Re}\xi (K_- -K_+) - i \text{Im}\xi  (K_+ + K_- + 2 K_0) + 2 i \text{Im}\xi \, K_0 } }
and apply the disentanglement procedure to split it into separate exponential operators 
\eqa{ e^{ \alpha (K_+ + K_- + 2 K_0) } \; e^{ \beta (K_- -K_+) } \; e^{ \gamma K_0 } = e^{ \alpha \o \sum_i x_i^2 } \; e^{ \beta (N/2 + \sum_i x_i \partial_{x_i} ) } \; e^{ \gamma H/(2\o) } }
whose action on a coordinate space wavefunction is transparent. 
We find
\eqa{ \alpha(\xi) & = -\frac i 2 \frac{\sin \phi \sinh 2 r}{ \cosh 2 r - \sinh 2 r \cos \phi } \| 
\beta(\xi) & = - \frac12 \log\( \cosh 2 r - \sinh 2 r \cos \phi \) \|
\gamma(\xi) & = \log\( \frac{ 1 - e^{-i\phi} \tanh r }{ 1 - e^{+i\phi} \tanh r } \) }
from which the evolved wavefunction turns out to be
\eqa[A5]{ \psi(\{x_i\},t) & =  \mathcal{N}_{N,\ell} \; (\o e^{2\beta+\gamma})^{N(1+\ell(N-1))/4}  e^{-\frac12 (e^{2\beta}-2\alpha)\o \sum_i x_i^2} \Delta(\{x_i\})^\ell \!
=  \mathcal{N}_{N,\ell} \; \[\o\frac{{1-|\eta(t)|^2}}{(1-\eta(t))^2}\]^{N [1+\ell (N-1)]/4}  e^{-\frac12 \( \frac{1+\eta(t)}{1-\eta(t)} \) \o \sum_i x_i^2}  \Delta(\{x_i\})^\ell 
}
with $  \eta(t) \equiv \kappa e^{-2i\o t} $, i.e. (\ref{5}) in the main text. 

\subsection{Time evolution of the density moments}

From the definition $\la x^{2n} \ra \equiv \int x^{2n} \varrho(x,t) dx $ of the density moments and our result (\ref{Omegat}) for the density $\varrho(x,t)$, we can derive their evolution in time
\eq{ \la x^{2n} \ra = \frac{1}{n+1}{2n\choose n} \(\frac{N\ell}{\Omega(t)} \)^n }
as well as the corresponding time averages
\eq[mom0]{ \overline{\la x^{2n} \ra } = \frac{1}{n+1}{2n\choose n}\Big{(}\frac{N\ell}{2\omega(1-\kappa^2)}\Big{)}^n B_n }
where $B_n \equiv ({\omega}/{2\pi})\int_0^{{2\pi}/{\omega}}(\kappa^2-2\kappa\cos 2\omega t+1)^n dt$. The $B_n$'s can be calculated explicitly 
\eq[Bn]{ B_n= \frac12 \[(1-\kappa)^{2n} \;  {}_2F_1\(\frac12,-n,1,\frac{-4\kappa}{(1-\kappa)^2}\) + (1+\kappa)^{2n} \; {}_2F_1\(\frac12,-n,1,\frac{4\kappa}{(1+\kappa)^2}\) \]}
where ${}_2F_1$ is the hypergeometric function. Note that $\overline{\la x^{2n}\ra}$ scales as $N^{2n}$ in the thermodynamic limit, as expected.

Another interesting observable is the correlation of the positions of different particles $\la x_1...x_{2m} \ra \equiv \int x_1... x_{2n} |\psi(\{x_i\},t)|^2 \, \prod_{i=1}^n dx_i$ which is \cite{Mehta04}
\eq{ \la x_1...x_{2n} \ra = \frac{(2n)!}{2^n n!} \(\frac{-\ell}{\Omega(t)}\)^{n}}
After time averaging we have
\eq{ \overline{\la x_1...x_{2n}\ra} =\(\frac{-\ell}{2\omega(1-\kappa^2)}\)^n B_n }
In contrast to the moments $\overline{\la x^{2n}\ra}$ the above correlations do not depend explicitly on $N$ therefore they are relatively unimportant in the thermodynamic limit.

\subsection{Derivation of the expansion of time averaged moments in terms of the IoM}

We will show that in the harmonic Calogero model the set of time averaged observables $\overline{\la x^{2n} \ra}$ can be expressed in terms of the integrals of motion $\la I_m \ra$ with $m=1,2,...,n$ and their products $\la \prod_{k} I_k \ra$ with $ \sum k \le n-1 $. It is instructive to start with the single harmonic oscillator case first, i.e. $\ell = 0$ and $N=1$.

The operator $x^{2n}$ is
\eq{ x^{2n} = \frac1{(2\o)^n} \( a + a^\dg \)^{2n} =  \frac1{(2\o)^n} \sum_{\{\sigma_r=\pm\}} \prod_{r=1}^{2n} a_{\sigma_r} }
where $a_+ \equiv a^\dg$ and $a_- \equiv a$. 
Evolving the operators in time (in the Heisenberg picture) according to $e^{iHt}a_i e^{-iHt} = a_i e^{-i\o t}$ and $e^{iHt}a_i^\dg e^{-iHt} = a_i^\dg e^{+i\o t}$ and taking the time average, we notice that only the ($(2n)!/(n!)^2$ in number) terms with equal number of $a$ and $a^\dg$ operators contribute
\eq[2]{ \overline{\la x^{2n} \ra} = \frac1{(2\o)^n} \sum_{\{\sigma_r=\pm\} \atop \sum \sigma_r = 0 } \left\la \prod_{r=1}^{2n} a_{\sigma_r} \right\ra_0 }
These terms can be re-ordered, using the CCR, as alternating sequences of $a^\dg$ and $a$, therefore giving
\eq{ \overline{\la x^{2n} \ra} = \frac1{(2\o)^n} \sum_{m=0}^n c_{n,m} \la (a^\dg a)^m \ra_0 }
for some appropriate combinatorial coefficients $c_{n,m}$. One way to derive the latter is by first normal-ordering the terms in (\ref{2}) and then re-expressing them in terms of $(a^\dg a)^m$. Both steps are known in the literature \cite{Wilc67}
\eq{ \overline{\la x^{2n} \ra} = \frac1{(2\o)^n} \sum_{l = 0}^n \frac{(2n)!}{2^{n-l} (n-l)! (l!)^2}  \la {a^\dg}^{l} a^{l} \ra_0 }
and
\eq{ {a^\dg}^{l} a^{l} = \sum_{m=0}^l s(l,m) \; (a^\dg a)^m }
where $s(l,m)$ are the Stirling numbers of the first kind, 
so that
\eq{ \overline{\la x^{2n} \ra}  = \frac1{(2\o)^n} \sum_{l = 0}^n \sum_{m=0}^l  \frac{(2n)!}{2^{n-l} (n-l)! (l!)^2}   \, s(l,m) \; \la (a^\dg a)^m \ra_0 }
from which we can read off the coefficients $c_{n,m}$ introduced above
\eq[cnm]{ c_{n,m} = \sum_{l=m}^{n}  \frac{(2n)!}{2^{n-l} (n-l)! (l!)^2}   \, s(l,m) }

We can easily generalise to the case of $N>1$ noninteracting bosons, for which the above results give
\eqa{ \overline{\la x^{2n}\ra} & \equiv \frac1N \sum_i \overline{\la x_i^{2n} \ra} \|
& = \frac1N \frac1{(2\o)^n} \sum_{m=0}^n c_{n,m} \sum_i \la (a_i^\dg a_i)^m \ra_0 \|
& = \frac1N \frac1{(2\o)^n} \sum_{m=0}^n c_{n,m} \la I_m \ra_0 }
Note that the knowledge of the values $\la I_m \ra$ is equivalent to the information of how many particles occupy each of the excitation levels of the trap, since from their generating function $\sum_j \exp{(is a^\dg_j a_j)} = \sum_{m=0}^\infty (is)^m I_m/m!$ one can derive the occupation number operator of the $n$-th level as $\sum_j \hat P_j(n) = \int_0^{2\pi} ds/(2\pi) \, e^{-ins} \sum_j \exp{(is \, a^\dg_j a_j)} $ where $\hat P_j(n) = |n\ra_j \, {}_j\la n|$ is the projector on the $n$-th level eigenstate of the $j$-th boson. In \cite{CSC13as} the set of IoM used in the construction of the GGE is the free fermion analogue of these occupation number operators of the trap levels. Therefore for the GGE in the interacting case, it is natural to choose the set (\ref{iom1}) which is the generalisation of the free boson or fermion sets.

The interacting case $\ell \neq 0$ is however different: due to the non-trivial commutation relations (\ref{ccr}), each time a swapping of two adjacent operators is needed in order to bring a term of (\ref{2}) into the required alternating form, additional lower order terms appear that involve a sum over the exchange operator $M_{ij}$. For example, let us consider the term $a_i^\dg a_i a_i a_i^\dg a_i^\dg a_i a_i^\dg a_i $ which needs only one swap of the 3rd and 4th operators in order to be brought into the right form. By doing so we get the extra term $\la a_i^\dg a_i (\sum_{j\neq i} M_{ij}) a_i^\dg a_i a_i^\dg a_i \ra $. Moving the exchange operator to the left or right i.e. to the bra or ket, on which its action gives simply 1, we change the particle index of the intermediate string of operators from $i$ to $j$ and as a result the final expression is 
$\sum_i \sum_{j\neq i} \la (a^\dg_i a_i) (a^\dg_j a_j)^2 \ra = \la I_1 I_2 \ra$, instead of $\sum_i \la (a^\dg_i a_i)^3 \ra = \la I_3 \ra $ as it would be if there was no exchange operator. In general we now have, apart from all previous terms, also all possible products $\la \prod_{s} I_{k_s} \ra$ where $\{k_s\}$ is any partition of the integer $m\le n-1$, thus leading to (\ref{taevo}) in the main text. The combinatorial coefficients $d_{n,m}$ and $e_{n,m,\{k_s\}}$ at leading order in $N$ for the first few moments are
\eqa[mom]{
\overline{\la x^2 \ra} & = \( 2 \la i_1 \ra_0 + \ell N \)/ {(2\o)} \|
\overline{\la x^4 \ra} & = \( 6 \la i_2 \ra_0 + 6 \ell N \la i_1 \ra_0 + 2 \ell^2 N^2 \) / {(2\o)^2}  \|
\overline{\la x^6 \ra} & = \( 20 \la i_3 \ra_0 + 20 \ell N \la i_2 \ra_0 + 20 \ell^2 N^2 \la i_1 \ra_0 + 5 \ell^3 N^3 + 10 \ell N \la i_1^2 \ra_0 \) / {(2\o)^3} \|
\overline{\la x^8 \ra} & = \( 70 \la i_4 \ra_0 + 70 \ell N \la i_3 \ra_0 + 70 \ell^2 N^2 \la i_2 \ra_0 + 70 \ell^3 N^3 \la i_1 \ra_0 + 14 \ell^4 N^ 4 + 70 \ell^2 N^2 \la i_1^2 \ra_0 + 70 \ell N \la i_1 i_2 \ra_0 \) / {(2\o)^4}
}
where $i_s \equiv I_s/N $. 

In order to calculate the values of the IoM $I_s$ in the initial state $|\Psi\ra$ we could use the expansion (\ref{psi})
of the latter in terms of the post-quench excitations on which the $I_s$ are diagonal. However the eigenvalues of $I_s$ on a general eigenstate are given only implicitly in the literature \cite{K96}. An alternative route is instead to express the $I_s$ in terms of the pre-quench creation-annihilation operators using the inverse of the Bogoliubov transformation (\ref{bog}), then normal-order the result using the pre-quench version of the CCR (\ref{ccr}) and act on $|\Psi\ra$. We can omit all terms that would be impossible to contribute to the final expressions for the expectation values at \emph{leading order} in $N$. We observe that the leading order in $N$ contribution to the expectation values comes from all terms that when normal-ordered give the maximal power $s$ of the operator $\sum_{j\neq i}M_{ij}$ (like e.g. the alternating term $(a_{0i}a_{0i}^\dg)^s |\Psi\ra$) which in turn when acting on $|\Psi\ra$ give simply $N^s$. In other words, we find that 
\eq{ (a_i^\dg a_i)^s |\Psi\ra = \ell^s N^s f_s(\kappa) |\Psi\ra + ... }
for a suitable function $f_s(\kappa)$. In the last equation the dots ``...'' denote pre-quench excitations that are accompanied by coefficients of lower order in $N$ and do not contribute to the leading order expressions when we apply the state $\la\Psi|$ on the left. According to the above, the leading order expectation values of the IoM on the initial state are
\eq{\la I_s\ra_0 = \ell^s N^{s+1} f_s(\kappa)}

Next we consider the products of IoM. If we start with a product of two IoM $I_s I_r$ we can readily see that, always to leading order in $N$
\eq{ I_s I_r |\Psi\ra= \sum_i (a_i^\dg a_i)^s \sum_j (a_j^\dg a_j)^r |\Psi\ra = \sum_i (a_i^\dg a_i)^s \({\textstyle\sum_j M_{ij}}\) (a_i^\dg a_i)^r |\Psi\ra }
i.e. in general, products of IoM can be constructed by introduction of $\sum_j M_{ij}$ between $(a_i^\dg a_i)^s$ operators. Therefore their study too reduces to a direct application of the previous observation and we have
\eq{\la\prod_k I_{s_k}\ra_0 = N^{\sum_k (s_k+1)} \prod_k f_{s_k}(\kappa)}
which proves the factorisation of the IoM in the initial state at leading order in $N$.

For the IoM and their products appearing in (\ref{mom}) we obtain explicitly
\eqa[iom2]{
\la i_1 \ra_0 & = \kappa^2 \frac{1}{1-\kappa^2} \, \ell N , \|
\la i_2 \ra_0 & = \kappa^2 \frac{1+\kappa^2}{(1-\kappa^2)^2} \, \ell^2 N^2 , \|
\la i_3 \ra_0 & = \kappa^2 \frac{1+3\kappa^2+\kappa^4}{(1-\kappa^2)^3} \, \ell^3 N^3 , \|
\la i_4 \ra_0 & = \kappa^2 \frac{1+6\kappa^2+6\kappa^4+\kappa^6}{(1-\kappa^2)^4} \, \ell^4 N^4 , \|
\la i_5 \ra_0 & = \kappa^2 \frac{1+10\kappa^2+20\kappa^4+10\kappa^6+\kappa^8}{(1-\kappa^2)^5} \, \ell^5 N^5 , \|
\la i_1^2 \ra_0 & = \kappa^4 \frac{1}{(1-\kappa^2)^2} \, \ell^2 N^2 , \|
\la i_1 i_2 \ra_0 & = \kappa^4 \frac{1+\kappa^2}{(1-\kappa^2)^3} \, \ell^3 N^3 .
}

As a verification of the consistency of our results, we compare the GGE values of the first four moments, as given by (\ref{mom}) along with (\ref{iom2}), with the actual time averaged values, as given by (\ref{mom0}) and  (\ref{Bn}) from which we have 
\eqa{
B_1&=1+\kappa^2,\|
B_2&=1+4\kappa^2+\kappa^4,\|
B_3&=1+9\kappa^2+9\kappa^4+\kappa^6 , \|
B_4&=1 + 16 \kappa^2 + 36 \kappa^4 + 16 \kappa^6 + \kappa^8.}
The agreement between these two independently derived results is indeed exact.


\begin{thebibliography}{99}


\bibitem{uc}
M.~Greiner, O.~Mandel, T.~W.~H\"ansch, and I.~Bloch,
Nature {\bf 419} 51 (2002).

\bibitem{kww-06}
T. Kinoshita, T. Wenger,  D. S. Weiss, 
 Nature {\bf 440}, 900 (2006).

\bibitem{tc-07}
S. Hofferberth, I. Lesanovsky, B. Fischer, T. Schumm, and J. Schmiedmayer,
Nature {\bf 449}, 324 (2007).

\bibitem{tetal-11}
S. Trotzky Y.-A. Chen, A. Flesch, I. P. McCulloch, U. Schollw\"ock,
J. Eisert, and I. Bloch, 
Nature Phys. {\bf 8}, 325 (2012). 

\bibitem{cetal-12}
M. Cheneau, P. Barmettler, D. Poletti, M. Endres, P. Schauss, T. Fukuhara, C. Gross, I. Bloch, C. Kollath, and S. Kuhr,
Nature {\bf 481}, 484 (2012).

\bibitem{getal-11}
M. Gring, M. Kuhnert, T. Langen, T. Kitagawa, B. Rauer, M. Schreitl, I. Mazets, D. A. Smith, E. Demler, and J. Schmiedmayer,
Science {\bf 337}, 1318 (2012).

\bibitem{shr-12}
U. Schneider, L. Hackerm\"uller, J. P. Ronzheimer, S. Will, S. Braun, T. Best, I. Bloch, E. Demler, S. Mandt, D. Rasch, and A. Rosch,
Nature Phys. {\bf 8}, 213 (2012).

\bibitem{rsb-13}
J. P. Ronzheimer, M. Schreiber, S. Braun, S. S. Hodgman, S. Langer, I. P. McCulloch, F. Heidrich-Meisner, I. Bloch, and U. Schneider, 
Phys. Rev. Lett. {\bf 110}, 205301 (2013).

\bibitem{revq}
A. Polkovnikov, K. Sengupta, A. Silva, and M. Vengalattore, 
Rev. Mod. Phys. {\bf 83}, 863 (2011).

\bibitem{gg} 
M. Rigol, V. Dunjko, V. Yurovsky,  and M. Olshanii,
Phys. Rev. Lett. {\bf 98}, 050405 (2007);
M. Rigol, V. Dunjko,  and M. Olshanii,
Nature {\bf 452}, 854 (2008). 

\bibitem{cc-06} P. Calabrese and  J. Cardy, 
Phys. Rev. Lett. {\bf 96}, 136801 (2006); 
J. Stat. Mech. ({\bf 2007}) P06008; J. Stat. Mech. ({\bf 2005}) P04010.

\bibitem{c-06}
M. A. Cazalilla, Phys. Rev. Lett. {\bf 97}, 156403 (2006); 
A. Iucci, and M. A. Cazalilla, New J. Phys. {\bf 12}, 055019 (2010).

\bibitem{c-06b}
A. Iucci, and M. A. Cazalilla, Phys. Rev. A {\bf 80}, 063619 (2009).

\bibitem{cdeo-08}
M. Cramer, C. M. Dawson, J. Eisert, and T. J. Osborne, 
Phys. Rev. Lett. {\bf 100}, 030602 (2008);
M. Cramer and J. Eisert,
New J. Phys. 12, 055020 (2010).

\bibitem{bs-08}
T. Barthel and U. Schollw\"ock, 
Phys. Rev. Lett. {\bf 100}, 100601 (2008).

\bibitem{r-09}
M. Rigol, Phys. Rev. Lett. {\bf 103}, 100403 (2009), Phys. Rev. A {\bf 80}, 053607 (2009)

\bibitem{CEF}
P. Calabrese, F.H.L. Essler and M. Fagotti, 
Phys. Rev. Lett. {\bf 106}, 227203 (2011);  J. Stat. Mech. ({\bf 2012}) P07016; ibid. ({\bf 2012}) P07022.

\bibitem{f-13}
M. Fagotti,  Phys. Rev. B {\bf 87}, 165106 (2013).

\bibitem{eef-12}
F. H. L. Essler, S. Evangelisti, and M. Fagotti,
Phys. Rev. Lett. {\bf 109}, 247206 (2012).

\bibitem{se-12}
D. Schuricht and F. H. L. Essler, J. Stat. Mech. ({\bf 2012}) P04017.

\bibitem{ccss-11}
T. Caneva, E. Canovi, D. Rossini, G. E. Santoro, and A. Silva,  J. Stat. Mech. ({\bf 2011}) P07015.

\bibitem{rs-12}
M. Rigol and M. Srednicki, Phys. Rev. Lett. {\bf 108}, 110601 (2012).

\bibitem{bdkm-11}
G. Biroli, C. Kollath, and A. M. L\"auchli,
Phys. Rev. Lett. {\bf 105}, 250401 (2010);
G. P. Brandino, A. De Luca, R.M. Konik, and G. Mussardo, 
Phys. Rev. B {\bf 85}, 214435 (2012).

\bibitem{fm-10}
D. Fioretto and G. Mussardo,
New J. Phys. {\bf 12}, 055015 (2010); 
B. Pozsgay, J. Stat. Mech. ({\bf 2011}) P01011.

\bibitem{mc-12a}
J. Mossel and J.-S. Caux, New J. Phys. {\bf 14},  075006 (2012).

\bibitem{ce-13}
J.-S. Caux and F. H. L. Essler, Phys. Rev. Lett. {\bf 110}, 257203 (2013).

\bibitem{fe-13}
M.  Fagotti and  F. H. L. Essler, Phys. Rev. B {\bf 87}, 245107 (2013).

\bibitem{CSC13a}
M. Collura, S. Sotiriadis and P. Calabrese, Phys. Rev. Lett. {\bf 110}, 245301 (2013).

\bibitem{gge-new}
B. Pozsgay, arXiv:1304.5374;
M.  Fagotti and  F. H. L. Essler, J. Stat. Mech. ({\bf2013}) P07012;
M. Kormos, A. Shashi, Y.-Z. Chou, J.-S. Caux, A. Imambekov, arXiv:1305.7202;
G. Mussardo, arXiv:1304.7599.

\bibitem{sfm-12}
S. Sotiriadis, D. Fioretto, and G. Mussardo,
J. Stat. Mech. ({\bf 2012}) P02017.

\bibitem{Calo}
F. Calogero, J. Math. Phys. {\bf 10}, 2191 (1969); {\bf 10} 2197 (1969); {\bf 12} 419 (1971).

\bibitem{Suth} 
B. Sutherland, Phys. Rev. A {\bf 4}, 2019 (1971); A {\bf 5}, 1372 (1972); Phys. Rev. Lett. {\bf 34}, 1083 (1975).

\bibitem{MC75}
J. Moser, Adv. Math. {\bf 16}, 197 (1975);
F. Calogero, Lett. Nuovo Cimento {\bf 13}, 411 (1975);
F.~Calogero, O.~Ragnisco and C.~Marchioro, Lett. Nuovo Cimento {\bf 13}, 383 (1975).

\bibitem{anyons}
J.~M.~Leinaas and J.~Myrheim, Phys. Rev. {\bf B} 37, 9286 (1988); A.~P.~Polychronakos, Nucl. Phys. {\bf B} 324, 597 (1989); Phys. Lett. B {\bf 264}, 362 (1991).

\bibitem{Poly92}
A.~P.~Polychronakos, Phys. Rev. Lett. {\bf 69}, 703 (1992); L.~Brink, T.~H.~Hansson, and M.~A.~Vasiliev, Phys. Lett. B {\bf 286} 109 (1992).

\bibitem{CSC13b}
M. Collura, S. Sotiriadis and P. Calabrese, 
J. Stat. Mech. ({\bf 2013}) P09025.

\bibitem{bch-11}
M. C. Banuls, J. I. Cirac, and M. B. Hastings,
Phys. Rev. Lett. {\bf 106}, 050405 (2011). 

\bibitem{CM11}
J.-S. Caux and J. Mossel, 
J. Stat. Mech. ({\bf 2011}) P02023.

\bibitem{Suth71}
B. Sutherland, J. Math. Phys. {\bf 12}, 246 (1971); {\bf 12}, 251 (1971)

\bibitem{Dunkl89}
C.~F.~Dunkl, Trans. Amer. Math. Soc. {\bf 311}, 167 (1989)

\bibitem{jap98}
H.~Ujino, A.~Nishino, M.~Wadati, Phys. Lett. A {\bf 249}, 459 (1998)

\bibitem{SGS13}
S.~Sotiriadis, A.~Gambassi and A.~Silva, Phys. Rev. E {\bf 87}, 052129 (2013)

\bibitem{Perel86}
A.~Perelomov, \emph{Generalized coherent states and their applications}, Springer-Verlag (1986)


\bibitem{SM}
See Supplemental Material.


\bibitem{Suth98}
B.~Sutherland, Phys. Rev. Lett. {\bf 80}, 3678 (1998)

\bibitem{GBD10}
V.~Gritsev, P.~Barmettler, E.~Demler, New J. Phys. {\bf 12}, 113005 (2010)

\bibitem{mg-05}
A. Minguzzi and D. M. Gangardt, Phys. Rev. Lett. {\bf 94}, 240404 (2005)

\bibitem{ciom} 
P. Calogero, C. Marchioro and O. Ragnisco, Lett. Nuovo Cim. {\bf 13}, 383 (1975); 
H. Ujino, K. Hikami and M. Wadati: J. Phys. Soc. Jpn. {\bf 61} 3425 (1992);
M. Wadati and H. Ujino, [arXiv:cond-mat/9706156]

\bibitem{mc-12b}
J. Mossel and J.-S. Caux, New J. Phys. {\bf 14},  075006 (2012).

\bibitem{yy69}
C. N. Yang and C. P. Yang, J. Math. Phys., {\bf 10}(7):1115-1122 (1969)

\bibitem{gp08}
D.M. Gangardt and M. Pustilnik, Phys. Rev. A {\bf 77}, 041604(R) (2008)

\bibitem{KE08}
M. Kollar and M. Eckstein, 
Phys. Rev. A {\bf 78}, 013626 (2008)


\end{thebibliography}

\begin{thebibliography}{99}

\bibitem{Wilc67}
R.~M.~Wilcox, J. Math. Phys. {\bf 8}, {962} (1967)

\bibitem{Mehta04}
See Eq.~(17.8.5) of L.~Mehta, \emph{Random matrices}, 
{Elsevier/Academic Press}, ({2004})

\bibitem{CSC13as}
M. Collura, S. Sotiriadis and P. Calabrese, Phys. Rev. Lett. {\bf 110}, 245301 (2013).

\bibitem{K96}
S.~Kakei,  J. Phys. A: Math. Gen. {\bf 29} L619 (1996)

\end{thebibliography}
\end{document}